\documentclass[journal=jacsat,manuscript=article,dvips]{achemso}
\usepackage{chemformula} 
\usepackage[T1]{fontenc} 
\usepackage{amsmath,amssymb}
\usepackage{bm}
\usepackage{colortbl}

\author{Hiroshi Takatsu}
\affiliation{Department of Energy and Hydrocarbon Chemistry, Graduate School of Engineering, Kyoto University, Kyoto 615-8510, Japan}
\author{Masayuki Ochi}
\affiliation{Department of Physics, Osaka University, Machikaneyama-cho, Toyonaka, Osaka 560-0043, Japan}
\author{Naoya Yamashina}
\affiliation{Department of Energy and Hydrocarbon Chemistry, Graduate School of Engineering, Kyoto University, Kyoto 615-8510, Japan}
\author{Morito Namba}
\affiliation{Department of Energy and Hydrocarbon Chemistry, Graduate School of Engineering, Kyoto University, Kyoto 615-8510, Japan}
\author{Kazuhiko Kuroki}
\affiliation{Department of Physics, Osaka University, Machikaneyama-cho, Toyonaka, Osaka 560-0043, Japan}
\author{Takahito Terashima}
\affiliation{Department of Physics, Graduate School of Science, Kyoto University, Kyoto 606-8502, Japan}
\author{Hiroshi Kageyama}
\affiliation{Department of Energy and Hydrocarbon Chemistry, Graduate School of Engineering, Kyoto University, Kyoto 615-8510, Japan}
\email{kage@scl.kyoto-u.ac.jp}

\title{%
Epitaxial stabilization of SrCu$_3$O$_4$ with infinite Cu$_{3/2}$O$_2$ layers
}

\begin{document}
\begin{abstract}
We report the {\color[rgb]{0,0,0} epitaxial thin film} synthesis of SrCu$_3$O$_4$ with infinitely stacked 
Cu$_3$O$_4$  {\color[rgb]{0,0,0} layers composed of edge-sharing CuO$_4$ square-planes}, using molecular beam epitaxy. 
Experimental and theoretical characterizations showed that this material is 
a metastable phase that can exist by applying tensile biaxial strain from 
the (001)-SrTiO$_3$ substrate. SrCu$_3$O$_4$ shows an insulating electrical resistivity in
accordance with the Cu$^{2+}$ valence state revealed X-ray photoelectron
spectroscopy. 
First-principles calculations also indicated that the unoccupied $d_{3z^2-r^2}$
band becomes substantially stabilized owing to the absence of apical anions, 
in contrast to $A_2$Cu$_3$O$_4$Cl$_2$ ($A = \rm{Sr}$, Ba) with an $A_2$Cl$_2$
block layer and therefore a {\it trans}-CuO$_4$Cl$_2$ octahedron. 
These results suggest that SrCu$_3$O$_4$ is a suitable parent material 
for electron-doped superconductivity based on the Cu$_3$O$_4$ plane.
\end{abstract}

\section{Introduction}
The discovery of superconductivity in La$_{2-x}$Ba$_x$CuO$_4$~\cite{Bednorz1986}
has led to intensive research to explore layered cupric compounds with high 
$T_{\rm c}$'s~\cite{ArmitageRMP2010,ChuPhysicaC2015,KeimerNature2015,TokuraNature1989-1,TokuraJJAP1990}. 
They universally possess two-dimensional (2D) CuO$_2$ planes with 
an apex-linked CuO$_4$ square-planar coordination, sandwiched by blocking 
layers (e.g., LaO layers in La$_2$CuO$_4$~\cite{TokuraJJAP1990} and BaO layers 
in YBa$_2$Cu$_3$O$_6$~\cite{WuPRL1987}). 
Superconductivity is observed when appropriate carriers are injected to 
the `parent' Cu$^{2+}$O$_2$ plane. Carrier doping can usually be done 
via aliovalent substitution or by creating defects in the blocking layers.
Despite extensive research, most cuprate superconductors are hole-doped, 
and only a handful electron-doped materials have been reported. 
In the electron-doped case, the absence of apical anions in 
the CuO$_2$ plane is considered as 
essential~\cite{ArmitageRMP2010,TokuraNature1989-1}. 
$A$CuO$_2$ ($A = $ alkaline earth metals)~\cite{SiegristNature1988,TakanoPhysicaC1989}, 
which consists of infinitely stacked CuO$_2$ planes, [CuO$_2$]$_\infty$,
separated only by $A$ cations, is therefore the ideal host system. 
High-pressure synthesis and thin film growth techniques allow 
the expansion of accessible compositional range. For example, thin film 
studies show superconductivity upon electron doping, namely in 
the infinite layer (IL) Sr$_{1-x}Ln_{x}$CuO$_2$ ($Ln = \rm{La}$, Nd) with 
$T_{\rm c} \sim 40$~K~\cite{SmithNature1991,ErPhysicaC1992}. 
More recently, the isoelectric nickel oxide film has been shown to 
exhibit superconductivity at about 10~K~\cite{LiNature2019}.
In parallel with compounds having CuO$_2$ planes, different structural 
motifs have been explored~\cite{Buschbaum1977,HiroiJSSC1991}. 
In particular, the Cu$_{3/2}$O$_2$ (or Cu$_3$O$_4$) plane in 
$A_2$Cu$_3$O$_4$Cl$_2$ ($A = \rm{Sr}$, Ba), characterized by 
a periodic insertion of $1/2$ Cu in the CuO$_2$ plane 
(Figure~\ref{fig.1}b, right), has attracted attention~\cite{Buschbaum1977}, 
since the Cu arrangement in the Cu$_3$O$_4$ plane is reminiscent of 
a line-centered square lattice known as the Lieb lattice~\cite{LiebPRL1989}, 
and electron-doped superconductivity is suggested. 
Unfortunately, the blocking $A_2$Cl$_2$ layer in $A_2$Cu$_3$O$_4$Cl$_2$ provides
apical (Cl) anions to Cu(1), resulting in {\it trans}-CuO$_4$Cl$_2$ octahedra 
(Figure~\ref{fig.1}a), which is thus unsuitable for 
electron-doped superconductivity. 
In fact, the absence of superconductivity in hole-doped 
Ba$_2$Cu$_3$O$_4$Cl$_2$ has been discussed in terms of hole localization at 
the apical ligand~\cite{Noro1990,Kato2000}. 
It is thus crucial to explore materials with the Cu$_3$O$_4$ plane 
without apical anions. 
This paper reports on the thin film growth of SrCu$_3$O$_4$, 
using molecular beam epitaxy (MBE). 
In contrast to $A_2$Cu$_3$O$_4$Cl$_2$, 
this compound epitaxially grown on a SrTiO$_3$ substrate is composed of 
infinite layers of Cu$_{3/2}$O$_2$ (Cu$_3$O$_4$) separated only by Sr cations.
First-principles calculations verify that the biaxial tensile strain from 
the substrate stabilize SrCu$_3$O$_4$ against competing phases of 
``SrCu$_2$O$_3$ + CuO''. Furthermore, the $d_{3z^2-r^2}$ orbitals 
become substantially stabilized, making SrCu$_3$O$_4$ a unique candidate 
for electron-doped superconductor based on the Cu$_3$O$_4$ lattice.

\section{Experiments and Calculations}
Target films were grown on (001)-SrTiO$_3$ single crystalline substrates 
using a custom-made reactive MBE system (EGL-1420-E2, Biemtron). 
Elemental Sr and Cu fluxes were simultaneously provided from 
conventional Knudsen cells typically with flux rates of 
0.05~\AA/s for Sr and 0.02~\AA/s for Cu, as determined by 
an INFICON quartz crystal microbalance system before the growth. 
These rates approximately correspond to the nominal composition of 
Sr:Cu $=$ 1:2. 
Ozone gas was supplied using a commercial ozonizer with a background 
pressure of $4\times10^{-6}$~Torr. After annealing the substrates at 
about 700~$^\circ$C for 2--4 hours in vacuum, 
the temperature, monitored by an optical pyrometer 
{\color[rgb]{0,0,0} with the wavelength of $\lambda = 8$--$13$~$\mu$m (IR-CAI3TS, CHINO)}, 
was set at 472~$^\circ$C.
The surface structure of the film and the substrate was monitored
{\it{in}-\it{situ}} by refection high-energy electron diffraction (RHEED) 
with an acceleration voltage of 20 keV.
X-ray diffraction (XRD) measurements after the growth were carried out 
at room temperature (RT) using a Rigaku SmartLab diffractometer equipped with 
a Cu K$\alpha_1$ monochromator. The scanning transmission electron microscopy
(STEM) observation and chemical composition analysis were conducted using 
a JEOL transmission electron microscope (JEM-ARM200F) at an operating 
voltage of 200 kV equipped with an energy dispersive X-ray spectrometer
(JED-2300T SDD). 
Incident electron beam along the stacking direction of the film was 
transmitted through a hole made by removing the substrate by Ar-ion beam milling.
The valence state of copper was investigated by means of X-ray photoelectron
spectroscopy (XPS) with Mg K$\alpha$ radiation ($h\nu = 1253.6$~eV; Ulvac-Phi Model5500). 
The binding energies ($E_{\rm B}$) of each XPS spectrum with potential 
extrinsic $E_{\rm B}$ shifts caused by the charging effect were calibrated 
with the adventitious C 1s peak at 284.5~eV. As a reference of divalent copper,
XPS spectra for powder samples of SrCuO$_2$ and SrCu$_2$O$_3$ prepared under 
the reported conditions~\cite{TakanoPhysicaC1989,HiroiJSSC1991} were collected.
The electrical resistivity $\rho$ was measured by using a standard four-probe
method. 
Magnetization measurements were performed with a commercial superconducting
quantum interference device (SQUID) magnetometer (magnetic property 
measurement system (MPMS), Quantum Design).
To examine structural stability and electronic properties, 
we performed first-principles calculations using the projector augmented wave
method~\cite{KressePRB1999} 
as implemented in the Vienna {\it ab initio} simulation package (VASP)~\cite{KressePRB1993,KressePRB1994,Kresse1996,KressePRB1996}.
We used the Perdew-Burke-Ernzerhof parametrization of the generalized gradient
approximation (PBE-GGA)~\cite{PerdewPRL1996} without the spin-orbit coupling.
For calculations of the total energy and the phonon dispersion, we optimized 
the crystal structure until the Hellmann-Feynman force acting on each atom
becomes less than 0.01~eV\ \AA$^{-1}$. 
The plane-wave cutoff energy was 550~eV, and $10\times 4\times 10$, 
$10\times 14\times 10$, and $10\times 10\times 10$ $k$-meshes were used for 
CuO, SrCu$_2$O$_3$, and SrCu$_3$O$_4$, respectively. 
To calculate the phonon dispersion, we assumed for simplicity a non-magnetic
state using a $3\times 3\times 3$ $q$-mesh, using the experimental lattice
constants of the thin film ($a = 5.42$~\AA\ and $c = 3.53$~\AA). 
The finite displacement method, as implemented in 
the Phonopy software~\cite{TogoSM2015}, was used. 
The (non-magnetic) electronic band dispersion of SrCu$_3$O$_4$ was 
obtained using the structural parameters obtained by our experiments.

\section{Result and discussion}
%
We initially investigated the crystal structure of 
the obtained film {\color[rgb]{0,0,0} on the SrTiO$_3$ substrate} using high-resolution XRD. 
The out-of-plane $\theta$--$2\theta$ pattern shows peaks 
around $2\theta=25^\circ$ and 52$^\circ$, suggesting the growth of 
well-oriented epitaxial films (Supporting Information, Figure~S1). 
The out-of-plane lattice constant was estimated to be 3.53~\AA, 
slightly larger than the bulk samples of IL-SrCuO$_2$ 
(3.432~\AA)~\cite{TakanoPhysicaC1989} and spin-ladder SrCu$_2$O$_3$ 
(3.495~\AA)~\cite{HiroiJSSC1991}, both composed of Cu$^{2+}$--O layers 
with Sr$^{2+}$ cations in between. 
Distinct fringes in the XRD profile (Figure~S1) indicate the atomic scale
smoothness of the interface between the substrate and the film. 
The spacing of fringes gave a film thickness of about 30~nm, 
in agreement with that estimated from the growth rate. 
The rocking curve of the peak around 25$^\circ$ gave a full 
width at half-maximum (FWHM) value of $\Delta\omega =0.09^\circ$, 
indicating excellent film quality.
{\color[rgb]{0,0,0} No target phase was observed when KTaO$_3$ and DyScO$_3$ substrates 
with larger strain (vs. SrTiO$_3$) were used under the same growth conditions 
(Supporting Information, Figure~S2).}
%

\begin{figure}[h]
\begin{center}
\includegraphics[width=0.75\textwidth]{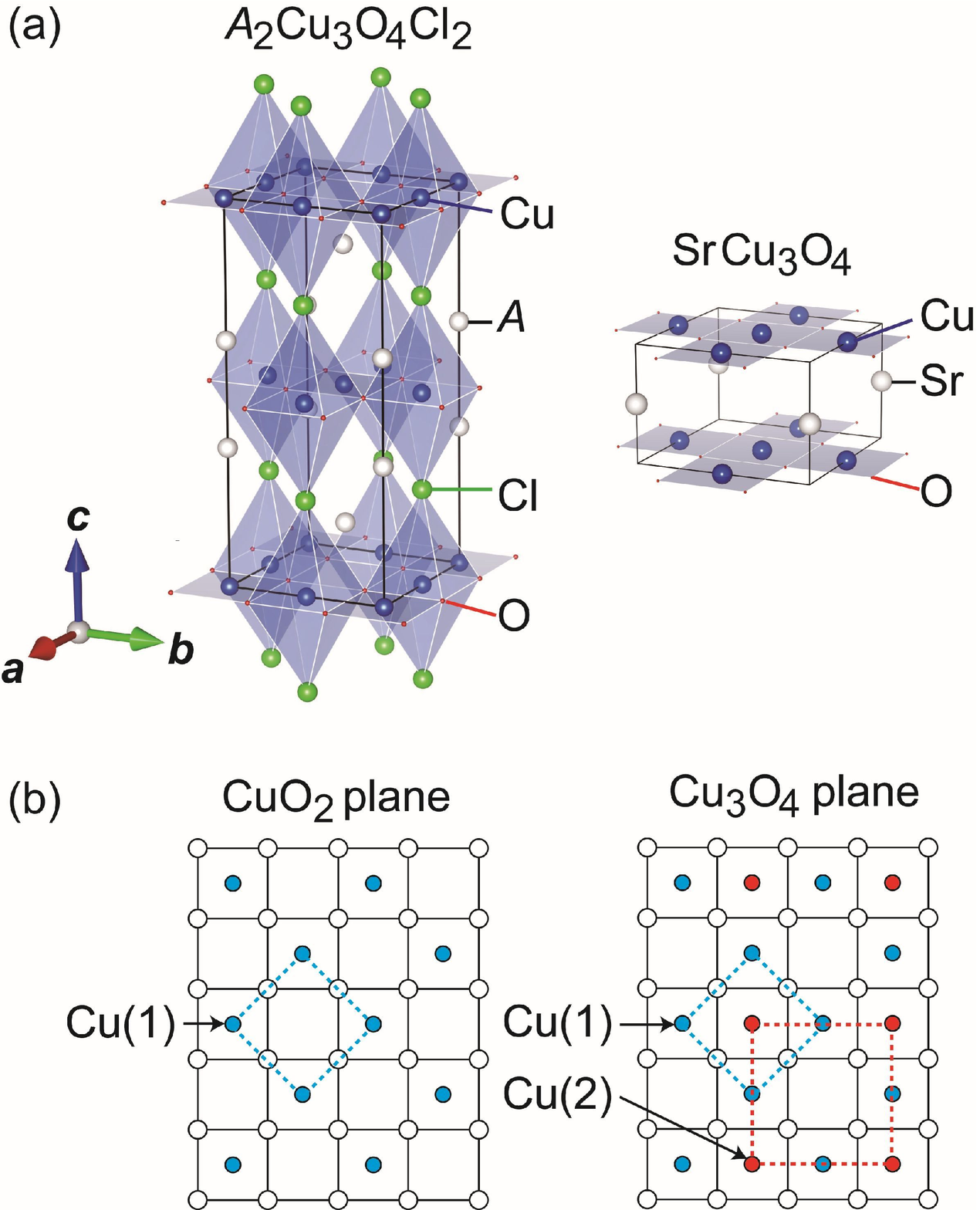}
\caption{
(a) Crystal structures of $A_2$Cu$_3$O$_4$Cl$_2$ ($A = \rm{Sr}$, Ba) (left) and
SrCu$_3$O$_4$ (right) with Cu$_3$O$_4$ (Cu$_{3/2}$O$_2$) planes. 
(b) 
The pristine CuO$_2$ plane with apex-linked CuO$_4$ units (left) 
and the Cu$_3$O$_4$ plane with edge-linked CuO$_4$ units (right) 
showing interpenetrating square lattices by Cu(1) and Cu(2) (dashed lines).
}
\label{fig.1}
\end{center}
\end{figure}

\begin{figure}[t]
\begin{center}
\includegraphics[width=0.85\textwidth]{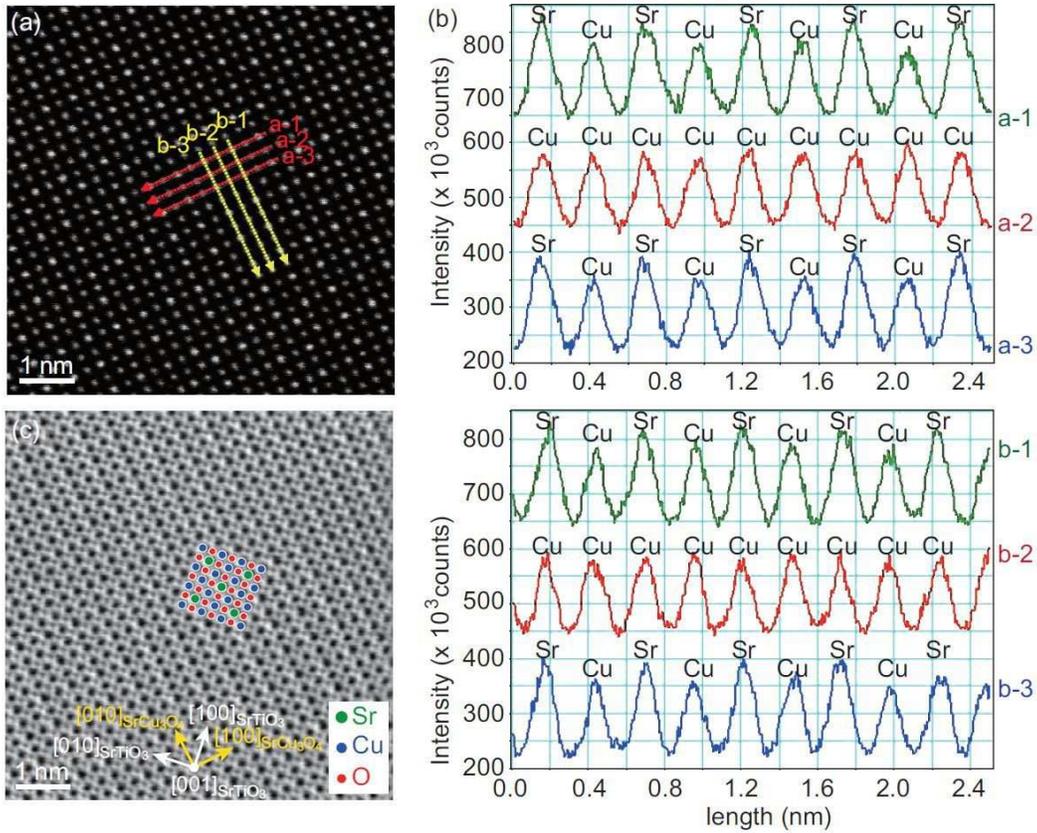}
\caption{
(a) High resolution HAADF-STEM image of the film with 30~nm 
thickness taken at RT. 
(b) Line scans along lines a-$i$ ($i=1,2,3$) (top) and 
those along lines b-$i$ ($i=1,2,3$) (bottom) shown in (a). 
(c) ABF-STEM image.
}
\label{fig.2}
\end{center}
\end{figure}

%
To clarify the in-plane structure of the film, we conducted STEM experiments.
Figure~\ref{fig.2}a shows a high-angle annular dark-field (HAADF) STEM image 
for a 30~nm thickness sample, highlighting the atomic position of Sr and Cu
corresponding, respectively, to bright and dark spots. 
The line scans in Figure~\ref{fig.2}b demonstrate the change in intensity $I$ 
at Sr and Cu positions. 
Since $I$ is proportional to the average atomic number $Z$ of the projected
atomic column and scales as $I\sim Z^n$  ($n =1.6$--1.9), these scans can
distinguish between Sr and Cu atoms. The observed pattern is different from
those expected from the CuO$_2$ plane (Figure~\ref{fig.1}b, left) and 
the BaCu$_3$O$_4$-type structure~\cite{Bertinotti1989}, but instead suggests 
the formation of Cu$_3$O$_4$ (Cu$_{3/2}$O$_2$) planes 
(Figure~\ref{fig.1}b, right), as found in $A_2$Cu$_3$O$_4$Cl$_2$, 
leading to SrCu$_3$O$_4$. 
Annular bright-field scanning transmission electron microscopy (ABF-STEM) is
also consistent with the formation of Cu$_3$O$_4$ (Cu$_{3/2}$O$_2$) planes
(Figure~\ref{fig.2}c). 
Note that EDX analysis exhibited a slightly smaller atomic ratio of
$\rm{Cu}/\rm{Sr} \simeq 2.5$, probably due to the influence of surface impurity
phases, as will be shown later. 
The corresponding electron diffraction along the $c$ axis 
(i.e., perpendicular to
the substrate) shows a four-fold symmetry with an in-plane lattice constant of 
$a = 5.42$~\AA\, (Supporting Information, Figure~S{\color[rgb]{0,0,0} 3}), 
which is consistent with the STEM image. 
The right panel of Figure~\ref{fig.1}a shows the most plausible structure of
SrCu$_3$O$_4$ with the $P4/mmm$ space group (No.~123), where Sr is at 
the 1b site, Cu at the 1c and 2f sites, and O at the 4j site 
(Supporting Information, Table~S1). 
The in-plane cell is related to the substrate (SrTiO$_3$ perovskite) 
by $\sqrt{2}a\times\sqrt{2}a$, as depicted in Figure~\ref{fig.2}c.
%

\begin{figure}[t]
\begin{center}
\includegraphics[width=0.80\textwidth]{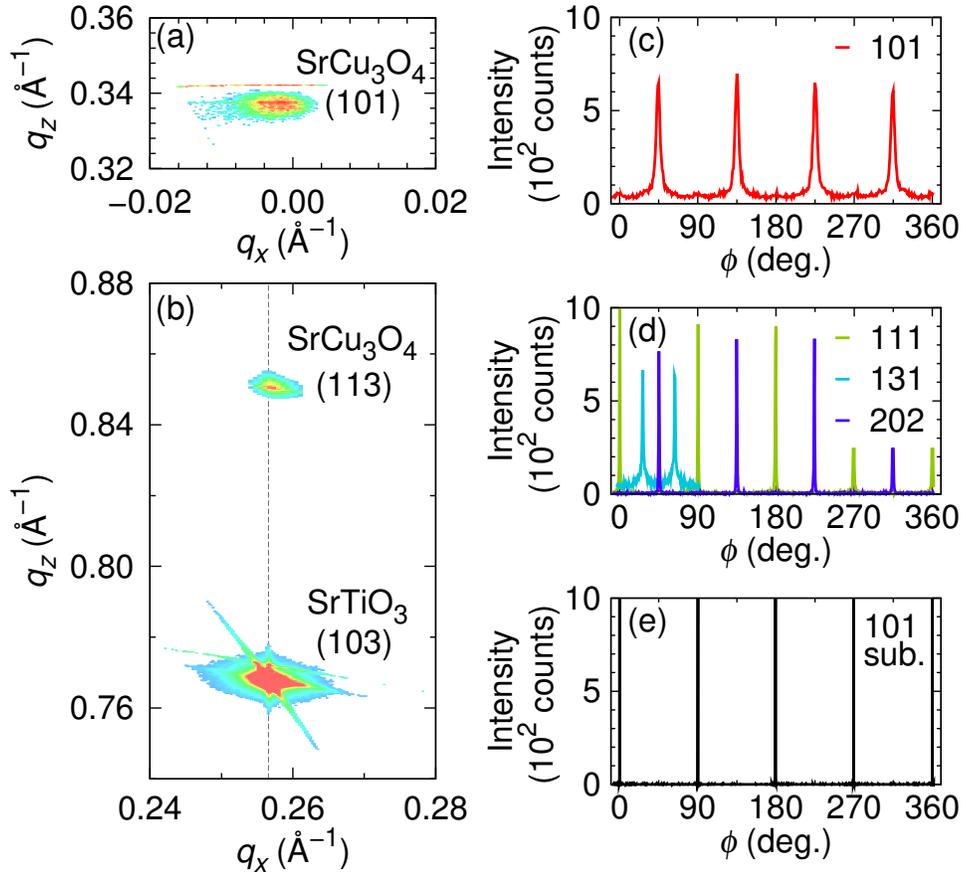}
\caption{
(a)--(b) X-ray reciprocal space mappings around 101 and 113 of 
the SrCu$_3$O$_4$ film. 
(c)--(e) $\phi$ scans of the 101 reflection of the SrTiO$_3$ 
substrate and representative peaks of the film.
}
\label{fig.3}
\end{center}
\end{figure}

%
The X-ray reciprocal space mapping (RSM) in Figure~\ref{fig.3}a shows 
a peak around $2\theta = 30.2^\circ$ and $\chi=33.1^\circ$, corresponding to 
the 101 refection of SrCu$_3$O$_4$. 
Known Sr-Cu-O compounds including Sr$_2$CuO$_3$~\cite{Liang1990}, SrCuO$_2$~\cite{TakanoPhysicaC1989}, 
Sr$_{14}$Cu$_{24}$O$_{41}$~\cite{McCarron1988}, SrCu$_2$O$_{2}$~\cite{Teske1970},
and Sr$_{\rm n-1}$Cu$_{\rm n+1}$O$_{2\rm n}$ (${\rm n} = 3,5$)~\cite{HiroiJSSC1991}
exhibit no reflection at this $2\theta$ angle. The azimuthal $\phi$ scan for 
the 101 reflection showing four peaks separated by 45$^\circ$ from 
the {101} peaks of the SrTiO$_3$ substrate (Figures~\ref{fig.3}c and \ref{fig.3}e) is consistent with 
the STEM image (Figures~\ref{fig.2}a and \ref{fig.2}c). 
Coherent growth of the film on the substrate is confirmed by the RSM around 
the 113 reflection (Figure~\ref{fig.3}b),
{\color[rgb]{0,0,0} and such coherent growth is reproduced in other films 
(Supporting Information, Figure~S4). The in-plane lattice constant estimated from RSM measurements, 
$a =  \sqrt{2}\times3.90 = 5.52$~\AA, was found to be 2\% more tensile-strained than that estimated 
from ED measurements ($a = 5.42$~\AA, Supporting Information, Table S2),
as a consequence of the removal of the SrTiO$_3$ substrate to observe 
the in-plane structure for ED measurements.}
Other reflections, such as 111, 131, and 202, are also compatible with 
the crystal symmetry of the SrCu$_3$O$_4$ structure (Figure~\ref{fig.3}d).
SrCu$_3$O$_4$ is the first experimental realization of Cu$_3$O$_4$ 
infinite layers (Figure~\ref{fig.1}a and \ref{fig.1}b, right). 
We note that the orthorhombic BaCu$_3$O$_4$ also has a layer of the Cu$_3$O$_4$
composition, but is composed of diamond chains (or diagonal ladders)~\cite{Bertinotti1989,RischauPRB2012}.
%

\begin{figure}[t]
\begin{center}
\includegraphics[width=0.75\textwidth]{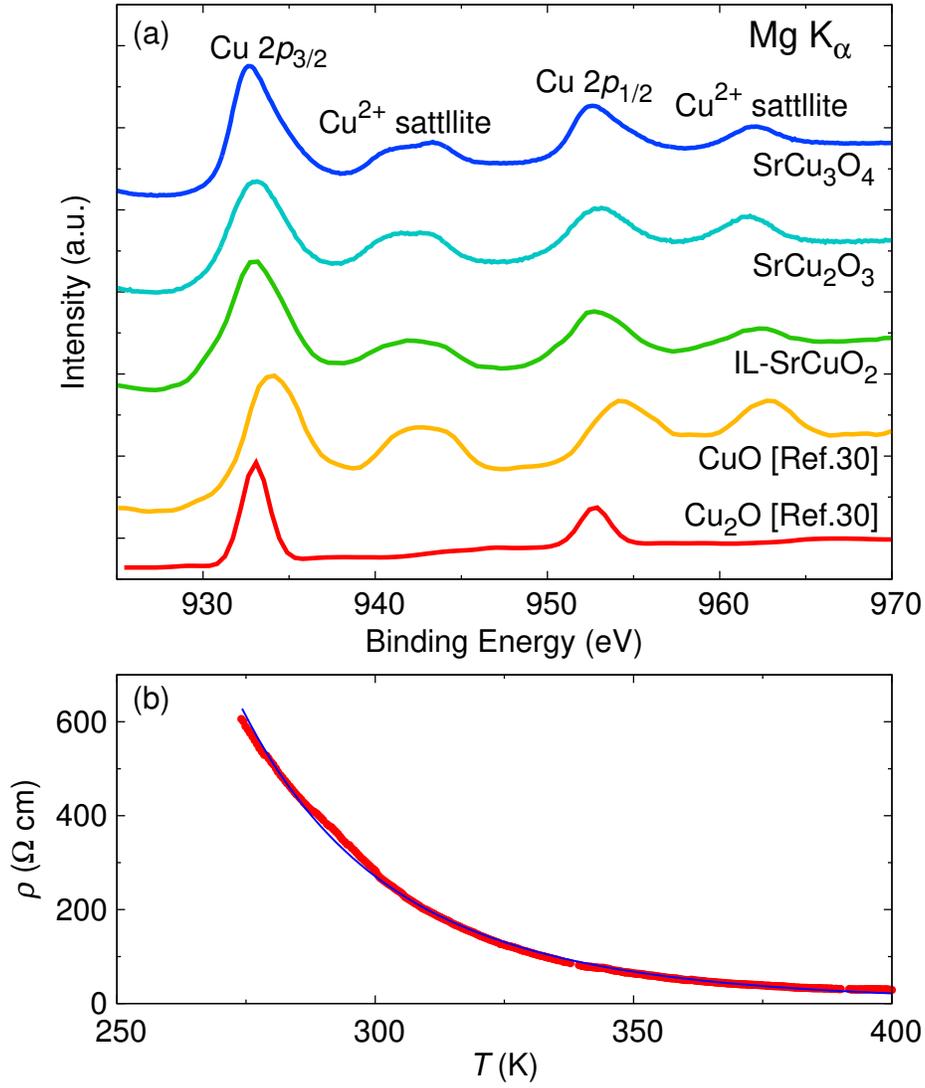}
\caption{
(a) XPS Cu 2$p$ core-level spectra of the SrCu$_3$O$_4$ film and several 
copper oxides. 
(b) Temperature dependence of electrical resistivity $\rho$ of 
the SrCu$_3$O$_4$ film.
The solid line represents the variable range hopping model for 2D (see text).
}
\label{fig.4}
\end{center}
\end{figure}

%
Using the lattice constants of $a=5.42$~\AA\, and $c=3.53$~\AA\, 
determined by XRD and STEM observations, we performed bond valence sum
calculations and obtained $+2.13$ for Sr and $+1.94$ for Cu, 
in accordance with the formal valences of Sr$^{2+}$ and Cu$^{2+}$. 
Figure~\ref{fig.4}a shows the 2$p$-core-level XPS spectrum of Cu in 
a SrCu$_3$O$_4$ film (5~nm thickness), plotted with SrCuO$_2$, SrCu$_2$O$_3$,
Cu$_2$O, and CuO~\cite{Hayez2004}. 
The spectrum has distinct satellite peaks separated by $\sim$9~eV, 
thus being reasonably ascribed to the divalent copper state.
%

%
Figure~\ref{fig.4}b shows the temperature dependence of the in-plane electrical
resistivity $\rho$ of the SrCu$_3$O$_4$ film with 30~nm thickness. 
It shows an insulating behavior in the temperature range from 400 to 270~K, 
but $\rho$ was too large to measure at lower temperatures. 
{\color[rgb]{0,0,0} Impurity scattering from defects and impurities will have a negligible effect on 
the resistivity in the temperature range currently measured~\cite{LeeRMP1985}.}
The $\rho$--$T$ curve can {\color[rgb]{0,0,0} thus} be fitted well by a variable range hopping in 
2D ($d = 2$), $\rho(T) = \rho_0 \exp(T_{0}/T)^{1/d+1}$, implying carrier 
hopping in the Cu$_3$O$_4$ plane. 
$A_2$Cu$_3$O$_4$Cl$_2$ with the Cu$_3$O$_4$
plane exhibits a canted antiferromagnetic phase transition of Cu(1) spins at 
$T_{\rm N} = 380$~K ($A = {\rm Sr}$)~\cite{ChouPRL1997} and 
$332$~K ($A = {\rm Ba}$)~\cite{ItoPRB1997}.
{\color[rgb]{0,0,0} No sign of antiferromagnetic phase transition is seen in 
the resistivity and magnetization data collected at $270~{\rm K} < T < 400 ~{\rm K}$ and $5~{\rm K} < T < 400~{\rm K}$, 
respectively, although SrCu$_3$O$_4$ is expected to be a Mott insulator. 
This is because it is difficult to probe antiferromagnetic order using thin films.}
%

\begin{figure}[t]
\begin{center}
\includegraphics[width=0.75\textwidth]{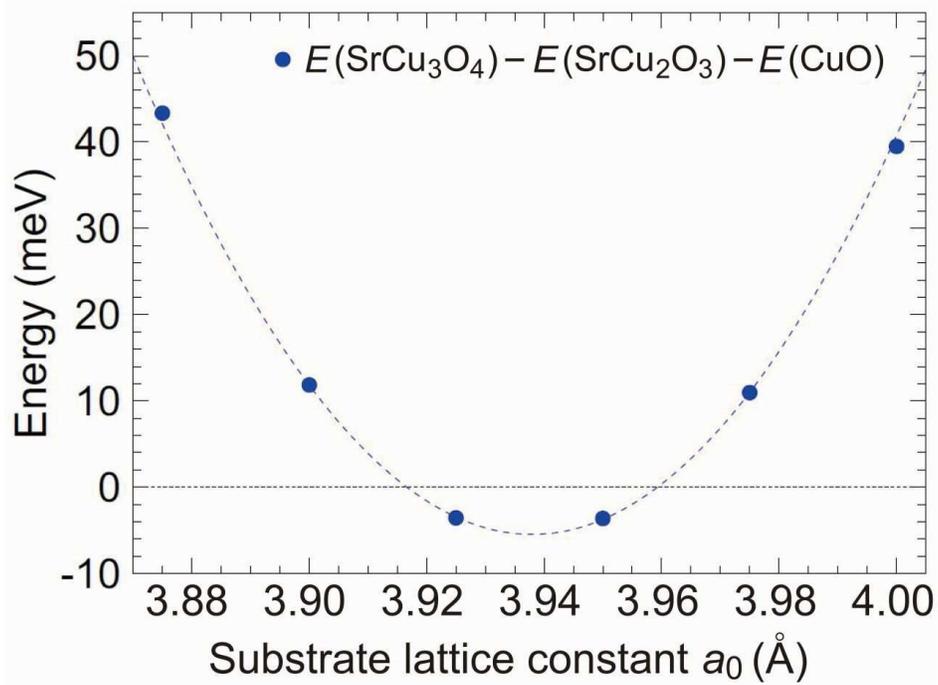}
\caption{
Energy difference between the calculated total energy of SrCu$_3$O$_4$ and 
that of SrCu$_2$O$_3$ $+$ CuO under the epitaxial strain, where $a_0$ is 
the in-plane lattice constant of the substrate. 
We assumed the fixed in-plane lattice constants $a = b = \sqrt{2}a_0$ 
for SrCu$_3$O$_4$ and $a = c = a_0$ for SrCu$_2$O$_3$. 
This line is the guide to the eye.
}
\label{fig.5}
\end{center}
\end{figure}

\begin{figure}[t]
\begin{center}
\includegraphics[width=0.75\textwidth]{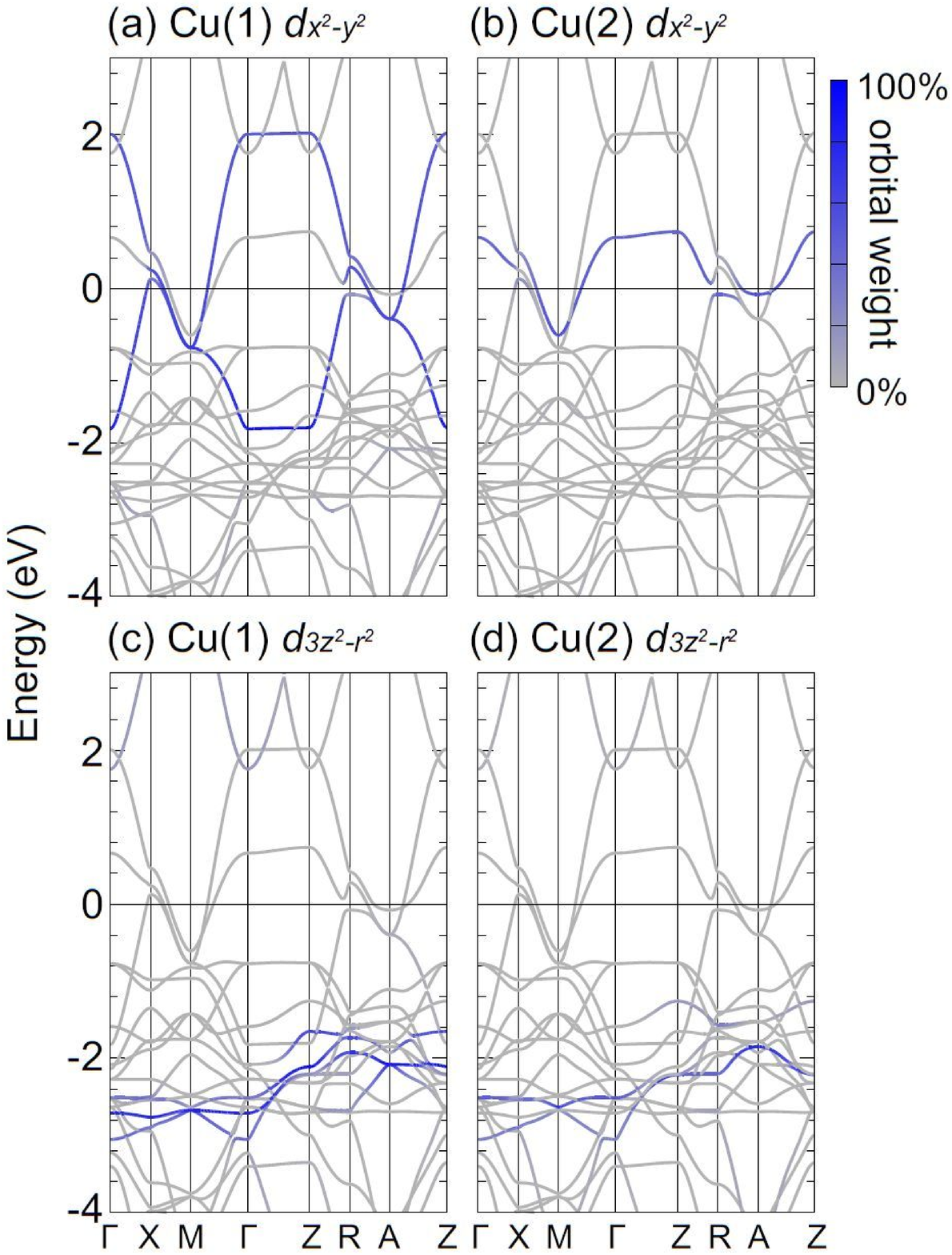}
\caption{
Calculated electronic band structure of SrCu$_3$O$_4$ with 
(a) Cu(1) $d_{x^2-y^2}$, (b) Cu(2) $d_{x^2-y^2}$, (c) Cu(1) $d_{3z^2-r^2}$, and
(d) Cu(2) $d_{3z^2-r^2}$ orbitals highlighted in blue. 
The Fermi energy is set at zero. 
Note that the notation of $d_{x^2-y^2}$ is used here in accordance with 
the usual manner of cuprates, while $d_{xy}$ is a more suitable notation to 
the present case since the $ab$ axes of SrCu$_3$O$_4$ are rotated by 45$^\circ$
from the $xy$ axes used in the calculations; i.e., the in-plane structure of
SrCu$_3$O$_4$ is expanded to $\sqrt{2}a \times \sqrt{2}a$ with respect to
IL-SrCuO$_2$. 
}
\label{fig.6}
\end{center}
\end{figure}

%
Let us discuss why the present MBE method has enabled the stabilization of
SrCu$_3$O$_4$. In general, MBE growth provides near thermal equilibrium
conditions, with much smaller kinetic energy than other growth techniques 
such as pulsed laser deposition and sputtering~\cite{SchlomJACS2008,OkaCrystEngComm2017}. 
For this reason, the synthesis of SrCu$_3$O$_4$ cannot be explained in terms of
kinetic trapping. 
On the other hand, epitaxial strain from the substrate is the most likely 
source for the SrCu$_3$O$_4$ phase formation~\cite{Gorbenko2002,Freund2003}.
This scenario is consistent with the observation that SrCu$_3$O$_4$ is 
formed only when the film thickness is less than 30~nm. 
For thicker films, impurity phases such as SrCu$_2$O$_3$ and CuO were 
observed in the STEM and RHEED images (Figure~S{\color[rgb]{0,0,0} 5}) and in the XPS data for 
the 30~nm film (Figure~S{\color[rgb]{0,0,0} 6}). 
In addition, finely tuned conditions are required {\color[rgb]{0,0,0} (Supporting Information, Table S3)}; 
SrCu$_3$O$_4$ can be prepared with a narrow growth window of temperature ($T = 472 \pm 1$~$^\circ$C)
and Cu/Sr ratio (nominal ratio of ${\rm Cu}/{\rm Sr} = 2 \pm 0.3$).
First-principles calculations were conducted to theoretically address 
the stability of SrCu$_3$O$_4$, in comparison with competing phases of 
SrCu$_2$O$_3$ and CuO. In an early stage of calculations, it was found 
that SrCu$_3$O$_4$ does not have an imaginary mode in the phonon dispersion 
when calculated using experimental lattice constants (Figure~S{\color[rgb]{0,0,0} 7}). 
However, the total energy of SrCu$_3$O$_4$ is higher than that of 
``SrCu$_2$O$_3$ + CuO'' by 0.19~eV per formula unit 
(here all the structural parameters were optimized), 
meaning that SrCu$_3$O$_4$ is dynamically stable 
{\color[rgb]{0,0,0} (or structurally stable as one of the possible structures 
in the Sr-Cu-O composition) but not the most thermodynamically stable}. 
In other words, this material cannot form in bulk owing to 
competing phases found in the Sr-Cu-O phase diagram~\cite{Thomas1992,HiroiPhysicaC1994,Moiseeva1998,KrockenbergerJAP2018}.
{\color[rgb]{0,0,0} It is interesting to compare the stability between SrCu$_3$O$_4$ and 
the competing SrCu$_2$O$_3$ from the structural perspective. SrCu$_3$O$_4$ consists only of 
edge-sharing CuO$_4$ square planes, 
whereas SrCu$_2$O$_3$ has both edge- and corner-sharing CuO$_4$ square planes~\cite{HiroiJSSC1991}.
Thus, the former structure appears to be more energetically unfavorable according to 
Pauling's third rule~\cite{Pauling_3rd_rule}. 
However, when tensile strain is applied to SrCu$_3$O$_4$,  
the Cu$_3$O$_4$ layer is stretched and the Cu-Cu repulsion between the edge-sharing 
CuO$_4$ square  planes is relaxed.}
We further compared the total energies of ``SrCu$_3$O$_4$'' and 
``SrCu$_2$O$_3$ $+$ CuO'' in the presence of epitaxial biaxial strain. 
To do this, in-plane lattice constants were fixed to the `normalized' substrate
lattice constant $a_0$ (where $a = b = \sqrt{2}a_0$ for SrCu$_3$O$_4$ and 
$a = c = a_0$ for SrCu$_2$O$_3$), and other structural parameters were optimized.
The strain effect for CuO was not taken into account since CuO was found as an
island on the film surface (Figure~S{\color[rgb]{0,0,0} 5}a). 
When biaxial strain is turned on (i.e., $a_0$ is varied), 
the relative energy of SrCu$_3$O$_4$, $E({\rm SrCu}_3{\rm O}_4) - E({\rm SrCu}_2{\rm O}_3) - E({\rm CuO})$, shows a parabolic shape with negative 
values for 3.91~\AA$<a_0<$3.96~\AA\,(Figure~\ref{fig.5}). 
The minimum is found at around 3.93~\AA, which is close to the lattice 
constant of SrTiO$_3$ (3.905~\AA) used as the substrate, 
suggesting that adequate tensile strain can stabilize the metastable phase of
SrCu$_3$O$_4$. 
The appearance of SrCu$_2$O$_3$ and CuO impurities for thicker films 
($> 30$~nm) would be due to some relaxation of the SrCu$_3$O$_4$ film from 
the substrate strain, as the film thickness increases. 
It is naturally considered that multiple domain formation can reduce the stress from the substrate~\cite{Gorbenko2002,LittlePRB1994}. 
In this case, when the film thickness exceeds the critical value, 
the epitaxially stable phase and bulk stable phase grow simultaneously~\cite{LittlePRB1994}. 
This is another strain relaxation mechanism different from 
the misfit dislocations that appear in the film-substrate 
interface~\cite{Freund2003,MatthewsJVST1975}.
{\color[rgb]{0,0,0} Cross-sectional TEM images will provide detailed features of 
the domain formations and other possible dislocations along the growth direction, 
which will be left for future studies.}
Finally, we discuss the difference in electronic structures between 
SrCu$_3$O$_4$ and $A_2$Cu$_3$O$_4$Cl$_2$. 
{\color[rgb]{0,0,0} Here, the main purpose of the band structure calculation is 
to see the band width of the $d_{x^2-y^2}$ orbital between Cu(1) and Cu(2), and 
the energy difference between $d_{x^2-y^2}$ and $d_{3z^2-r^2}$ orbitals. 
We thus performed the calculation assuming a non-magnetic and metallic state 
because otherwise the band width and the energy difference are strongly affected by 
the size of exchange splitting and become spin-dependent.}
The calculated band dispersion for ``non-magnetic'' SrCu$_3$O$_4$ near 
the Fermi energy (Figure 6a, 6b) show that the wide band from 
Cu(1) $d_{x^2-y^2}$ orbitals and the narrow band from Cu(2) $d_{x^2-y^2}$
orbitals are nearly decoupled, 
as found in $A_2$Cu$_3$O$_4$Cl$_2$~\cite{RosnerPRB1998}. 
Furthermore, the $d_{3z^2-r^2}$ bands of SrCu$_3$O$_4$ are substantially
stabilized (Figure~\ref{fig.6}c, \ref{fig.6}d) owing to 
the square-planar coordination, in contrast to $A_2$Cu$_3$O$_4$Cl$_2$ containing
chloride anions at the apical Cu(1) site. 
This means that SrCu$_3$O$_4$ is an ideal 2D system with 
half-filled  $d_{x^2-y^2}$ orbitals on the Cu$_3$O$_4$ lattice, 
where electron-doped superconductivity may occur because the absence of 
apical anions is known to be crucial for electron-doped superconductivity in 
the CuO$_2$ case~\cite{ArmitageRMP2010,TokuraNature1989-1}. 
If electron doping is possible in SrCu$_3$O$_4$, the absence of apical anions 
at the Cu(1) site is expected to prevent carrier localization, as observed in the CuO$_2$ system~\cite{Noro1990,Kato2000}. 
The ``inserted'' Cu(2) may bring about an exotic superconducting state or
magnetic ground state, for example, by ferromagnetic fluctuations introduced through 90$^{\circ}$-type Cu(1)--O--Cu(2) interaction. 
Electron doping by chemical substitution, as demonstrated in La- or
Nd-substituted IL-SrCuO$_2$~\cite{ArmitageRMP2010,SmithNature1991,ErPhysicaC1992}
and electric field gating~\cite{UenoJPSJ2014} in SrCu$_3$O$_4$ films, 
would be an interesting subject in the future.

\section{Conclusion}
We have succeeded in synthesizing SrCu$_3$O$_4$ thin films on a SrTiO$_3$
substrate using molecular beam epitaxy. 
SrCu$_3$O$_4$ consists of infinitely stacked 
Cu$_{3/2}$O$_2$ (Cu$_3$O$_4$) planes separated only by Sr cations, 
in contrast to $A_2$Cu$_3$O$_4$Cl$_2$ ($A = {\rm Sr}$, Ba) with 
the $A_2$Cl$_2$ block layer. 
Experimental and theoretical investigations revealed that this material is 
a metastable phase that can exist by applying suitable tensile biaxial 
strain from the substrate. 
Thus stabilized SrCu$_3$O$_4$ shows an insulating behavior, 
as expected from the formal Cu$^{2+}$ valence. 
The band structure with substantially stabilized (unoccupied) $d_{3z^2-r^2}$
orbitals (vs $A_2$Cu$_3$O$_4$Cl$_2$) owing to the absence of apical anions
suggests that this material is a suitable parent material for electron-doped
superconductivity based on the Cu$_3$O$_4$ lattice.

\section{Associated content}
\begin{suppinfo}
Additional information on results of X-ray diffraction, electron diffraction,
HAADF-STEM, XPS, and calculations of phonon dispersion of SrCu$_3$O$_4$.
\end{suppinfo}
%

\begin{acknowledgement}
This work was supported by CREST (JPMJCR1421) and JSPS KAKENHI 
Grants (No.~17H04849, No.~17H05481, No.~JP18H01860, and No.~JP18K13470, No.19H04697, No.20H00384).
\end{acknowledgement}
\bibliography{reference}

\newpage

\clearpage
\section{TOC graphic and synopsis}

We prepared epitaxial SrCu$_3$O$_4$ thin films with infinitely stacked Cu$_3$O$_4$ layers 
composed of edge-sharing CuO$_4$ square-planes, using molecular beam epitaxy. 
Experimental and theoretical characterizations revealed that 
SrCu$_3$O$_4$ can exist under tensile biaxial strain from the (001)-SrTiO$_3$ substrate. 
This material is a suitable system for 
electron doped superconductivity based on the Cu$_3$O$_4$ plane.  

\begin{figure}[h]
\begin{center}
\includegraphics[width=8.5cm]{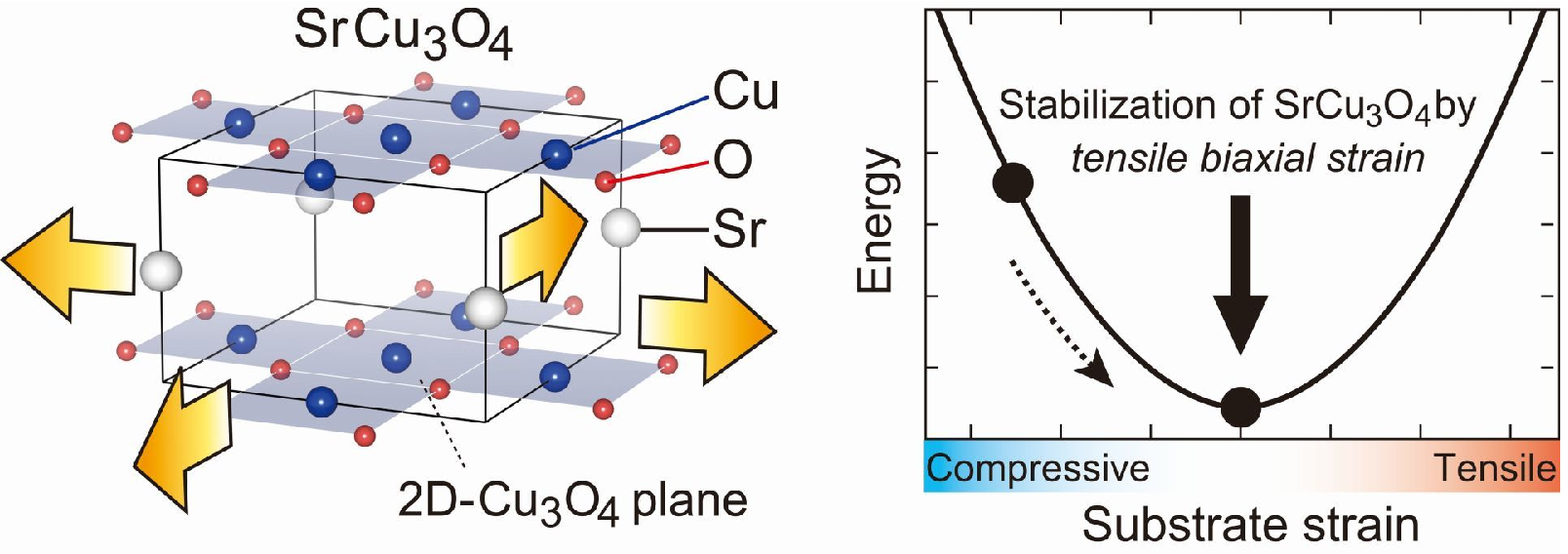}\\
\label{TOC}
For Table of Contents Only
\end{center}
\end{figure}

\clearpage

\begin{center}
\textbf{\Large Supporting Information of ``Epitaxial stabilization of SrCu$_3$O$_4$ with infinite Cu$_{3/2}$O$_2$ layers''}
\end{center}
\setcounter{equation}{0}
\setcounter{figure}{0}
\setcounter{table}{0}
\setcounter{page}{1}
\renewcommand{\theequation}{S\arabic{equation}}
\renewcommand{\thefigure}{S\arabic{figure}}
\renewcommand{\bibnumfmt}[1]{[S#1]}
\renewcommand{\citenumfont}[1]{S#1}

\section*{Abstract}
In this supporting information, we present X-ray diffraction (XRD), 
experimental and calculation results of electron diffraction (ED) for 
SrCu$_3$O$_4$ films grown on the (001)-SrTiO$_3$ substrate. 
We also show results of high-angle annular dark-field scanning transmission
electron microscopy (HAADF-STEM) for thick (30~nm) films, and 
X-ray photoelectron spectroscopy (XPS) for the thin (5~nm) and thick (30~nm)
films. In addition, we present the first principles calculation of the phonon
dispersion of SrCu$_3$O$_4$ with the experimental lattice constants 
($a = 5.42$~\AA\, and $c = 3.53$~\AA).

\section{{\color[rgb]{0,0,0}XRD analysis of the SrCu$_3$O$_4$ film on the SrTiO$_3$ substrate}}
\begin{figure}[h]
\begin{center}
\includegraphics[width=0.80\textwidth]{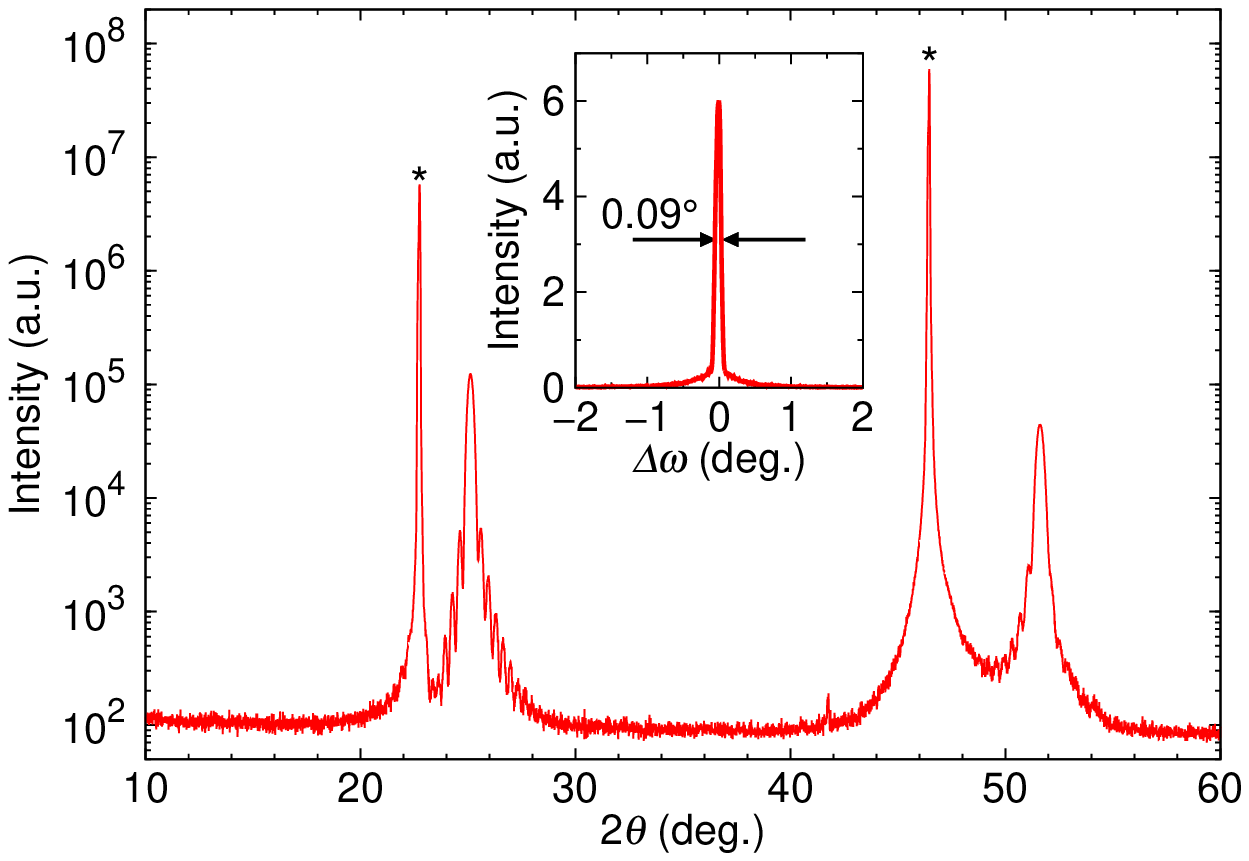}
\caption{
Out-of-plane $\theta$--$2\theta$ XRD patterns of films with a 30~nm thickness.
SrTiO$_3$ substrate peaks are marked with asterisks. 
Inset shows rocking curve of the peak around 25$^{\circ}$.
}
\label{fig.s1}
\end{center}
\end{figure}

Figure~\ref{fig.s1} shows the out-of-plane $\theta$--$2\theta$ XRD pattern for 
the film on the (001)-oriented SrTiO$_3$ substrate grown in the present 
optimal growth condition ($T = 472^{\circ}$C, nominal ratio of 
${\rm Cu}/{\rm Sr} = 2$, and O$_3$ gas flow with a background pressure of 
$4\times10^{-6}$~Torr). 
Sharp peaks with distinct fringes were observed around $2\theta=25^{\circ}$ and
$52^{\circ}$, suggesting the successful growth of a well-oriented and excellent
quality film, although a tiny unknown peak is present at around $42^{\circ}$. 
A film thickness is estimated from the spacing of fringes to be about 30~nm,
which agrees with the estimation from the growth rate. 
The inset shows the rocking curve of the peak around 25$^{\circ}$, 
showing a full width at half-maximum of $\Delta\omega =0.09^\circ$. 
In Table~\ref{table.s1}, we summarized structural parameters of SrCu$_3$O$_4$.
%
\begin{table}[h]
\begin{center}
 \caption{Structural parameters for the thin film of SrCu$_3$O$_4$.
 The space group: $P4/mmm$ (No.~123); lattice constants: 
 $a = 5.42$~\AA, and $c = 3.53$~\AA.
 }
 \begin{tabular*}{0.80\textwidth}{@{\extracolsep{\fill}}ccccc}
  \hline\hline
   Atom     & Site   & x        & y    & z     \rule{0mm}{4mm}   \\ \hline
   \,Sr     & 1b     & 0        & 0    & 1/2   \rule{0mm}{4mm}   \\
   \,Cu     & 1c     & 1/2      & 1/2  & 0     \rule{0mm}{3.5mm} \\
   \,Cu     & 2f     & 0        & 1/2  & 0     \rule{0mm}{3.5mm} \\
   \,O      & 4j     & 0.25     & 0.25 & 0     \rule{0mm}{3.5mm} \\ 
  \hline\hline
 \end{tabular*}
 \label{table.s1}
\end{center}
\end{table}

{\color[rgb]{0,0,0}
\section{Growth of films on other substrates}
%
\begin{figure}[t]
\begin{center}
\includegraphics[width=0.80\textwidth]{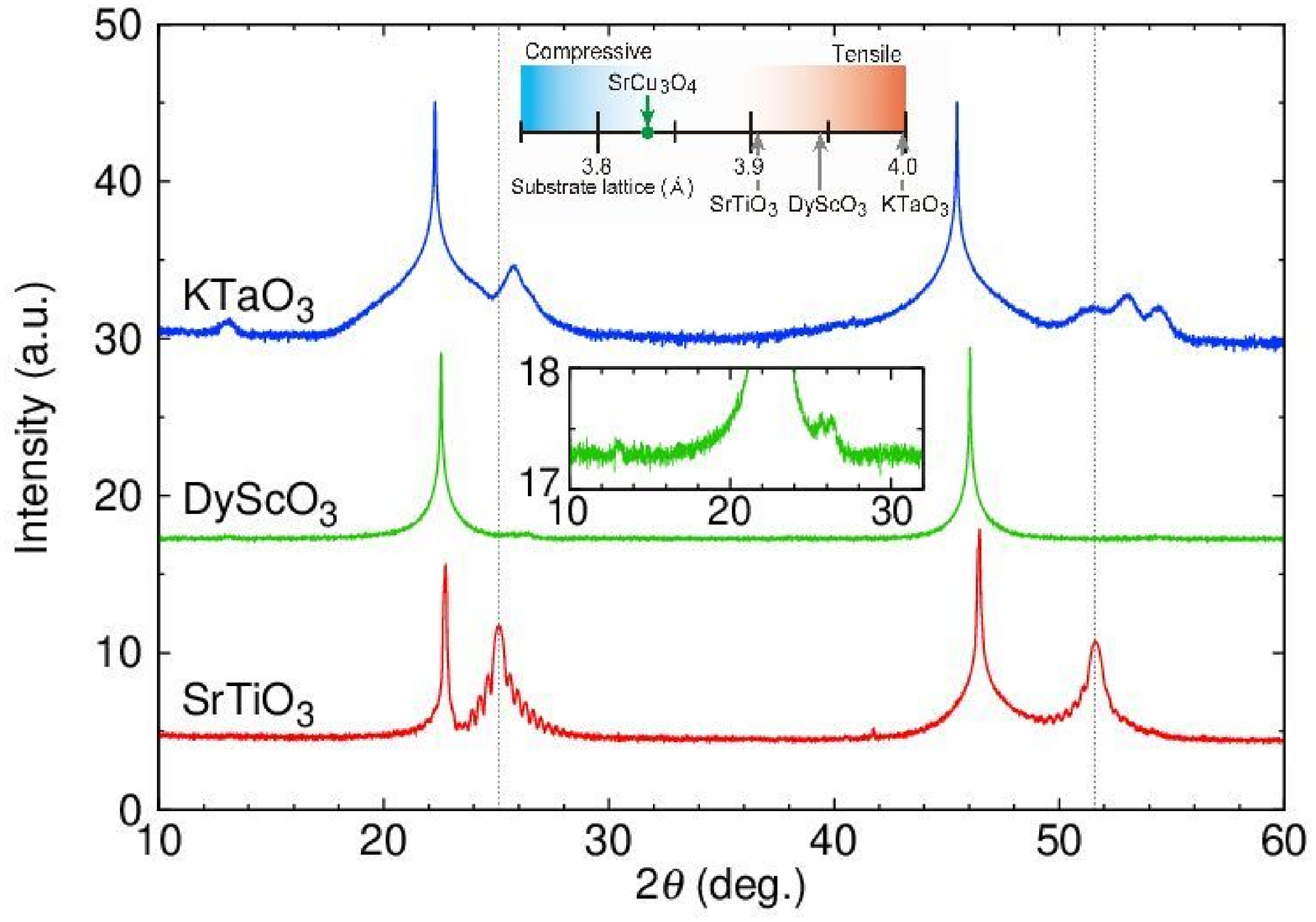}
\caption{
{\color[rgb]{0,0,0}
Out-of-plane $\theta$--$2\theta$ XRD patterns of films on the SrTiO$_3$, 
DyScO$_3$, and KTaO$_3$ substrates. 
The upper inset shows the relation between the substrate lattice constant 
and the in-plane lattice constant of SrCu$_3$O$_4$ divided by $\sqrt{2}$. 
The lower inset shows the enlarged XRD data of the film on 
the DyScO$_3$ substrate. 
Vertical dot lines indicate peak positions of SrCu$_3$O$_4$.
}
}
\label{fig.s2}
\end{center}
\end{figure}
%
We also performed experiments using other substrates than SrTiO$_3$, such as 
the (001)-oriented KTaO$_3$ and (110)-oriented DyScO$_3$ substrates under 
the present optimal growth condition for the growth of SrCu$_3$O$_4$ films 
on the SrTiO$_3$ substrate. 
For these KTaO$_3$ and DyScO$_3$ substrates, one can expect larger tensile
strain than the SrTiO$_3$ substrate for SrCu$_3$O$_4$ 
(see the inset of Figure~\ref{fig.s2})~\cite{SchlomMRS2014}. 
However, we did not obtain the single phase of SrCu$_3$O$_4$ on these 
substrates, respectively. 
Using the KTaO$_3$ substrate, we obtained a film containing three phases 
such as $c$-axis oriented SrCu$_2$O$_3$, $a$-axis oriented Sr$_{14}$Cu$_{24}$O$_{41}$, 
and possibly $c$-axis oriented SrCu$_3$O$_4$ (Figure~\ref{fig.s2}). 
On the other hand, using the DyScO$_3$ substrate, we obtained 
a film containing two phases such as $c$-axis oriented SrCu$_2$O$_3$, and 
$a$-axis oriented Sr$_{14}$Cu$_{24}$O$_{41}$, although XRD intensities of 
these phases are very small (Figure~\ref{fig.s2}). 
Since DyScO$_3$ has a pseudo-cubic structure~\cite{VelickovZK2007}, 
it is also possible that SrCu$_3$O$_4$ could not be obtained due to 
anisotropic strain. 
The obtained phases here are determined by XRD peak positions of
Sr-Cu-O compounds~\cite{Thomas1992,HiroiPhysicaC1994,Moiseeva1998}. 
We discuss more details of the strain effect for the synthesis of 
SrCu$_3$O$_4$ in the main text of this paper.
}

\newpage
\section{Experiments and calculations for ED of the SrCu$_3$O$_4$ film}
\begin{figure}[t]
\begin{center}
\includegraphics[width=0.80\textwidth]{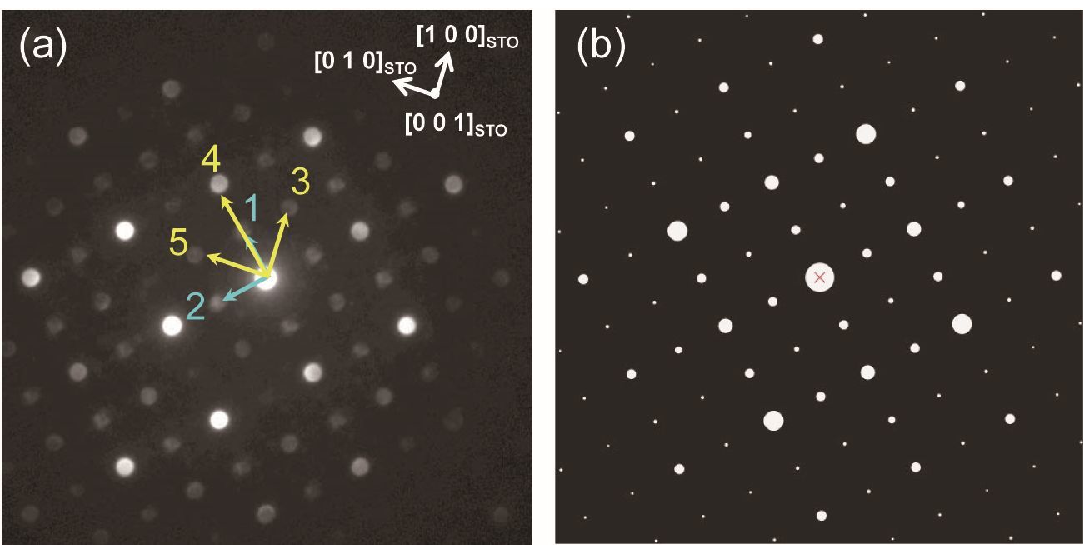}
\caption{
(a) Electron diffraction pattern of the film taken at room temperature along
[001]. (b) Calculated pattern of electron diffraction, assuming a tetragonal
structure with the in-plane lattice constant of 5.42~\AA.
}
\label{fig.s3}
\end{center}
\end{figure}

In Figure~\ref{fig.s3}a, we present experimental ED patterns for 
the same SrCu$_3$O$_4$ film as shown in Figure~2 of the main text. 
One can clearly see that the in-plane structure has a four-fold symmetry. 
The in-plane lattice constant is estimated as $a = 5.42$~\AA. 
This observation is in good agreement with the calculation of the ED pattern
(Figure~\ref{fig.s3}b), assuming the tetragonal unit cell of the given
structure of SrCu$_3$O$_4$ (see the right panel of Figure 1a of 
the main text and Table~\ref{table.s1}). 
We summarized observed and calculated $d_{hkl}$ parameters in Table~\ref{table.s2}.
%
%
%
\begin{table}[t]
\begin{center}
 \caption{Observed and calculated $d_{hkl}$ for representative ED spots. 
Numbers correspond to spot numbers in Figure~\ref{fig.s2}a.
 }
 \begin{tabular*}{0.80\textwidth}{@{\extracolsep{\fill}}cccc}
  \hline\hline
   No.     & measured $d_{hkl}$ (nm)   & calculated $d_{hkl}$ (nm)  & hkl           \rule{0mm}{4mm}   \\ \hline
   \,1     & 0.541                     & 0.542                      & 100           \rule{0mm}{4mm}   \\
   \,2     & 0.543                     & 0.542                      & 010           \rule{0mm}{3.5mm} \\
   \,3     & 0.384                     & 0.383                      & 1$\bar{1}$0   \rule{0mm}{3.5mm} \\
   \,4     & 0.271                     & 0.271                      & 200           \rule{0mm}{3.5mm} \\ 
   \,5     & 0.384                     & 0.383                      & 110           \rule{0mm}{3.5mm} \\ 
  \hline\hline
 \end{tabular*}
 \label{table.s2}
\end{center}
\end{table}

\newpage
\section{{\color[rgb]{0,0,0}Growth condition and reproducibility of the SrCu$_3$O$_4$ growth}}
\begin{table}[htb]
\begin{center}
\caption{{\color[rgb]{0,0,0}
Summary of the growth conditions (temperature and background pressure) for
SrCu$_3$O$_4$ films, with a nominal Cu/Sr ratio of 2. 
Double circles indicate that the XRD data showed the expected peaks and fringes.
A single circle means that the film possesses the expected peaks, but no fringe
is present. 
Triangles indicate the inclusion of multiple phases, including SrCu$_3$O$_4$,
and crosses indicate the absence of peaks from SrCu$_3$O$_4$.
}
}
 \begin{tabular}{|c|c|c|c|c|}
  \hline
   $T$(K)/$P$(Torr)   & \hspace{5mm}$3.0\times10^{-6}$\hspace{5mm}   & \hspace{5mm}$3.5\times10^{-6}$\hspace{5mm} & \hspace{5mm}$4.0\times10^{-6}$\hspace{5mm} & \hspace{5mm}$5.0\times10^{-6}$\hspace{5mm} \rule{0mm}{4mm}   \\ \hline
   \,440              & -         & $\times$             & -                        & -                   \rule{0mm}{5.5mm} \\ \hline
   \,450              & -         & $\times$             & -                        & -                   \rule{0mm}{5.5mm} \\ \hline
   \,455-458          & -         & $\times$             & $\times$                 & -                   \rule{0mm}{5.5mm} \\ \hline
   \,460              & -         & $\times$             & $\times$                 & -                   \rule{0mm}{5.5mm} \\ \hline
   \,461-463          & $\times$  & $\times$             & $\times$                 & -                   \rule{0mm}{5.5mm} \\ \hline
   \,465              & -         & -                    & \small{$\triangle$}      & -                   \rule{0mm}{5.5mm} \\ \hline
   \,470-471          & -         & \small{$\triangle$}  & \small{$\triangle$}      & -                   \rule{0mm}{5.5mm} \\ \hline
   \,472-473          & -         & \bm{$\circledcirc$}  & \bm{$\circledcirc$}      & \LARGE{$\circ$}     \rule{0mm}{5.5mm} \\ \hline
   \,473-474          & -         & \LARGE{$\circ$}      & \LARGE{$\circ$}          & -                   \rule{0mm}{5.5mm} \\ \hline
   \,475              & -         & -                    & \small{$\triangle$}      & -                   \rule{0mm}{5.5mm} \\ \hline
   \,475-477          & -         & -                    & \small{$\triangle$}      & -                   \rule{0mm}{5.5mm} \\ \hline
   \,480-482          & -         & -                    & $\times$                 & -                   \rule{0mm}{5.5mm} \\ \hline
 \end{tabular}
 \label{table.s3}
\end{center}
\end{table}
%
{\color[rgb]{0,0,0}
For the determination of optimal growth conditions, 
we performed experiments by changing temperature and the flow rate of O$_3$ gas. 
In Table~\ref{table.s3}, we summarized the results of 
the synthesis of SrCu$_3$O$_4$ on the SrTiO$_3$ substrate at various temperatures and background pressures 
at a Cu/Sr ratio of 2. Here, the quality of the film was checked by XRD peak positions, 
fringes in the out-of-plane $\theta$--$2\theta$ XRD scan, impurities and multiple phases. 
In the present study, the optimal growth condition for 
the SrTiO$_3$ substrate was $472\pm1^{\circ}$C and $4\times10^{-6}$~Torr 
at a nominal Cu/Sr ratio of 2. 
Note that in the same temperature and pressure conditions, the single phase of
SrCu$_3$O$_4$ was not obtained at different Cu/Sr ratios such as 
$\rm{Cu}/\rm{Sr} = 1$, 1.5, 1.7 and 2.3.
}

%
\begin{figure}[t]
\begin{center}
\includegraphics[width=0.95\textwidth]{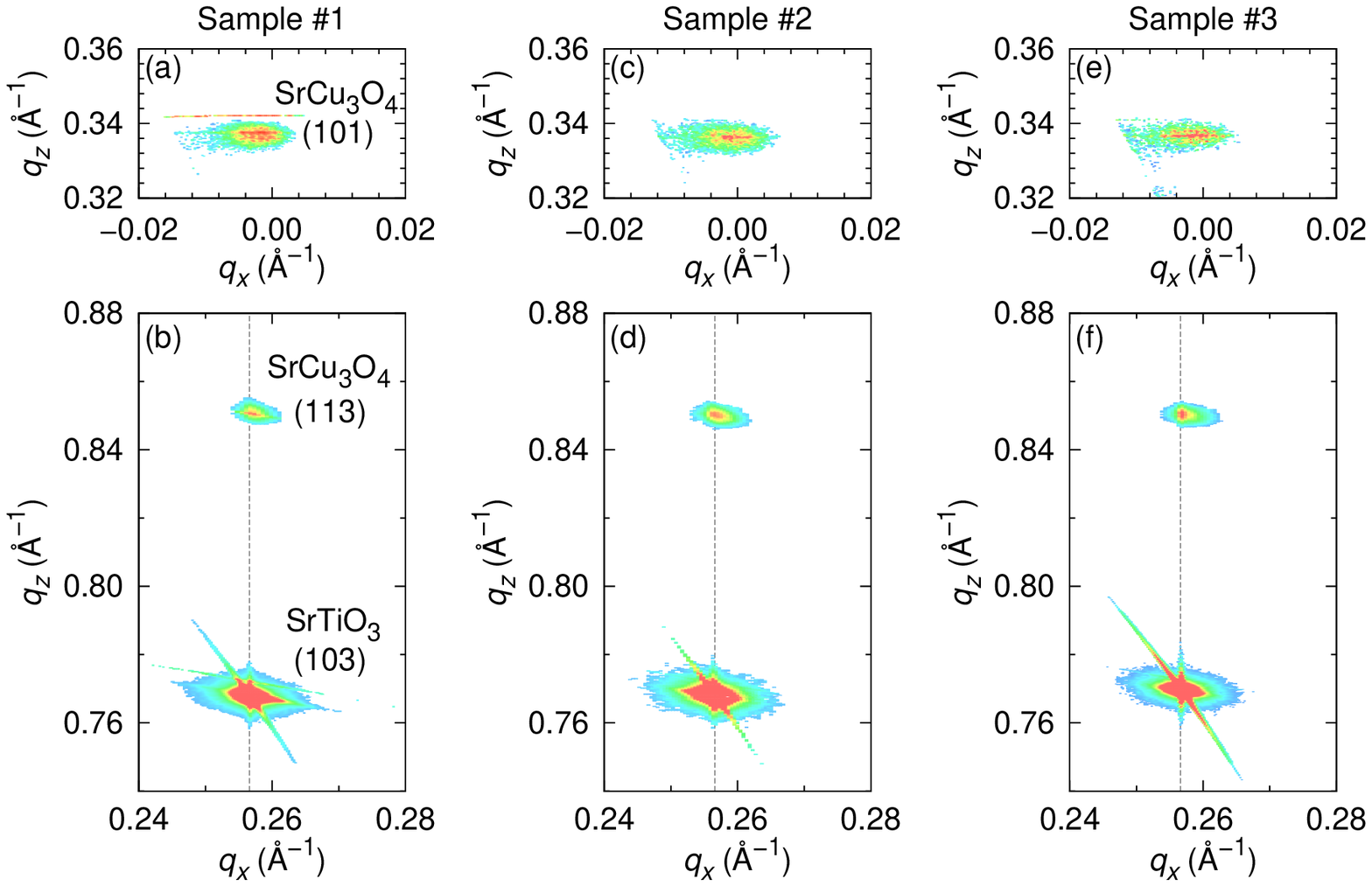}
\caption{
{\color[rgb]{0,0,0}
X-ray reciprocal space mappings around (a), (c), (e) 101 and (b), (d), (f) 113
for three different films of SrCu$_3$O$_4$. The thicknesses of these films is
almost the same ($\sim30$~nm). The coherent and reproducible growth of these films is confirmed.
}
}
\label{fig.s4}
\end{center}
\end{figure}
%
%
We observed the generation of non-crystalline phases and/or
Sr$_{14}$Cu$_{24}$O$_{41}$ with some impurity phases when growth conditions
are deviated from the optimal condition. 
For example, in the case of ${\rm Cu}/{\rm Sr} = 1$, SrCuO$_2$ with 
the infinite CuO$_2$ layer structure was obtained. 
The excess Sr condition for the growth of SrCu$_3$O$_4$ implies that 
the initially formed SrCu$_2$O$_3$ may be reconstructed to SrCu$_3$O$_4$ 
owing to the substrate strain as discussed in the main text, 
or re-evaporation of Sr from the film surface may take place. 

{\color[rgb]{0,0,0}
The synthesis of SrCu$_3$O$_4$ is highly reproducible; 
using the present optimal growth condition and the (001)-SrTiO$_3$ substrate, 
we have succeeded in synthesizing SrCu$_3$O$_4$ in more than 10 times. 
It is noted that, to maintain the stability of temperature and monitoring condition, 
we synthesized the films at a fixed position without rotating the substrates.
Almost the same coherent feature of obtained films was confirmed in reciprocal
space mapping of XRD measurements (Figure~\ref{fig.s4}).
}

\section{Multiple domain formation in thick films of SrCu$_3$O$_4$}
%
\begin{figure}[t]
\begin{center}
\includegraphics[width=0.80\textwidth]{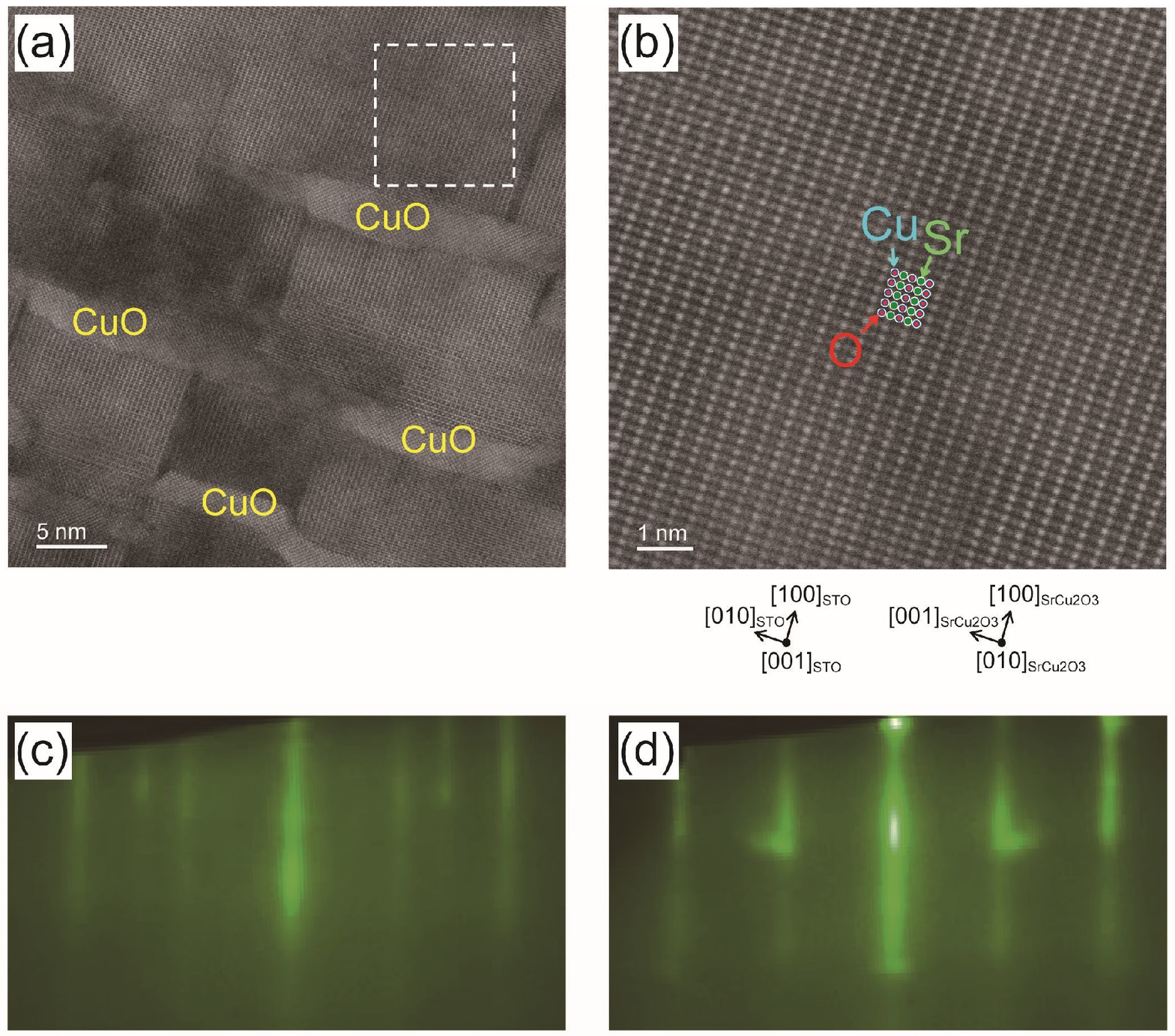}
\caption{
(a) HAADF-STEM image of an island of secondary phases in a thick film with 30~nm.
 Secondary phases were identified as b-axis oriented SrCu$_2$O$_3$ and CuO. 
(b) Magnified view for the white square region in (a). 
(c) RHEED images along the [100] direction of the SrTiO$_3$ substrate for 
the samples with 5~nm thickness and (d) 30~nm thickness.
}
\label{fig.s5}
\end{center}
\end{figure}
%
\begin{figure}[t]
\begin{center}
\includegraphics[width=0.80\textwidth]{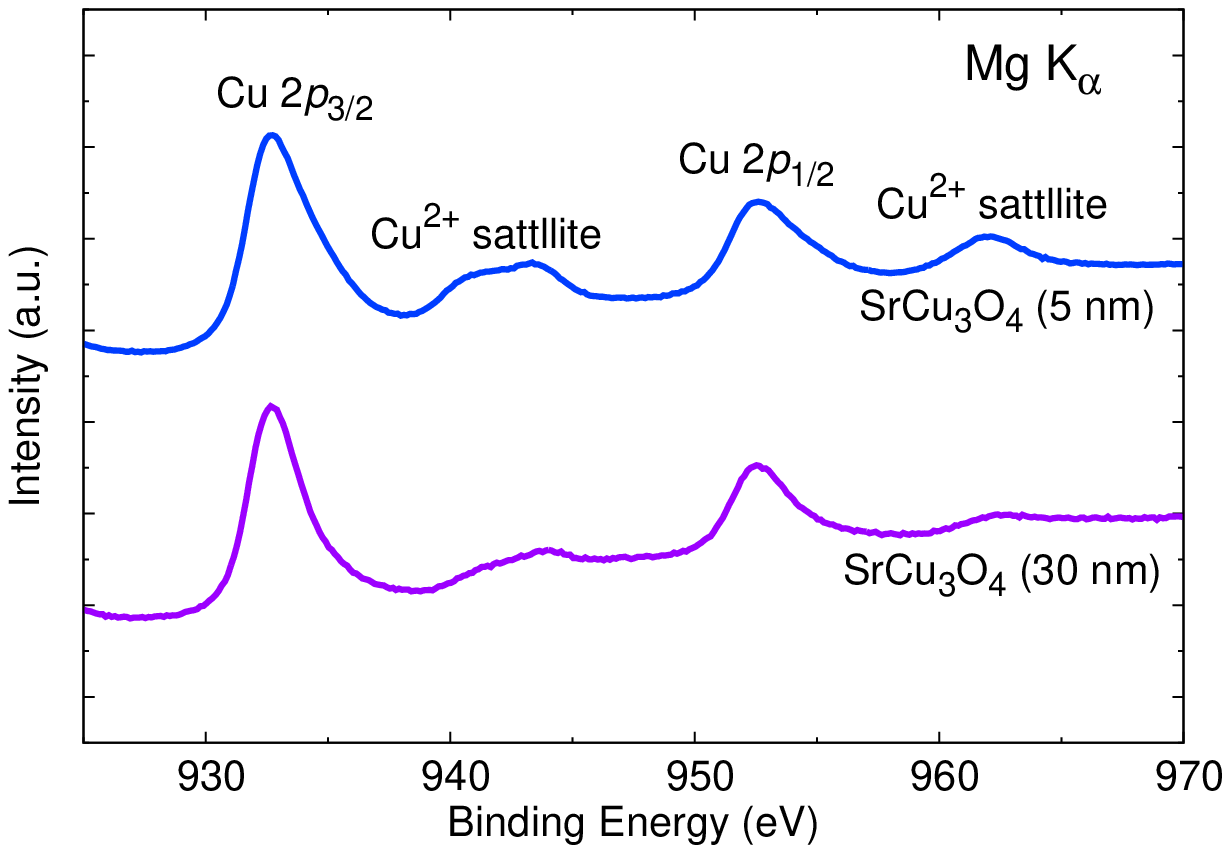}
\caption{
Cu 2$p$ core level XPS spectra of SrCu$_3$O$_4$ films with 5~nm and 30~nm thickness.
}
\label{fig.s6}
\end{center}
\end{figure}
%
A small portion of competing phases of SrCu$_2$O$_3$ and CuO was found in 
HAADF-STEM data for thicker films ($> 30$~nm), as shown in 
Figures~\ref{fig.s5}a and \ref{fig.s5}b. 
Competing impurity phases were also detected as spot-like features in 
the RHEED pattern. A typical RHHED image for the film with a 5~nm thickness
shows streaky patterns (Figure~\ref{fig.s5}c), suggesting an extremely flat
surface of the SrCu$_3$O$_4$ film. However, further growth under 
the same condition resulted in spot-like features along with the original
streaks (Figure~\ref{fig.s5}d), suggesting the formation of islands of 
extra phases that coexist with the epitaxially grown phase.

Figure~\ref{fig.s6} shows the 2$p$ core level of Cu for SrCu$_3$O$_4$ films 
on the (001)-SrTiO$_3$ substrate. The film with 5~nm thickness is 
the same sample as shown in Figure~4a of the main text. 
Here we also show the result of a thicker film with 30~nm. 
For the thin film with 5~nm, we observed distinct satellite peaks, 
separated by $\sim9$~eV, which are ascribed to the divalent copper state~\cite{Hayez2004}. 
However, intensities of the satellites decreased in the thicker film 
with 30~nm. This result is probably due to impurity phases in the surface 
state, which appear in thicker films than 30~nm.
A phase mixture of SrCu$_3$O$_4$ and other phases can be understood as 
a consequence of multiple domain formation that reduces the stress from 
the substrate~\cite{LittlePRB1994,Gorbenko2002}.
This result implies that the epitaxially stable phase (SrCu$_3$O$_4$) and
secondary phases (SrCu$_2$O$_3$ and CuO) are both thermodynamically 
stable~\cite{LittlePRB1994}, however, the former phase cannot appear in 
bulk owing to competing secondary phases that are more stable in bulk. 
The multiple domain formation is another strain relaxation mechanism 
different from misfit dislocation~\cite{MatthewsJVST1975,Freund2003}.

\clearpage

\section{Calculated phonon dispersion of SrCu$_3$O$_4$}
\begin{figure}[h]
\begin{center}
\includegraphics[width=0.80\textwidth]{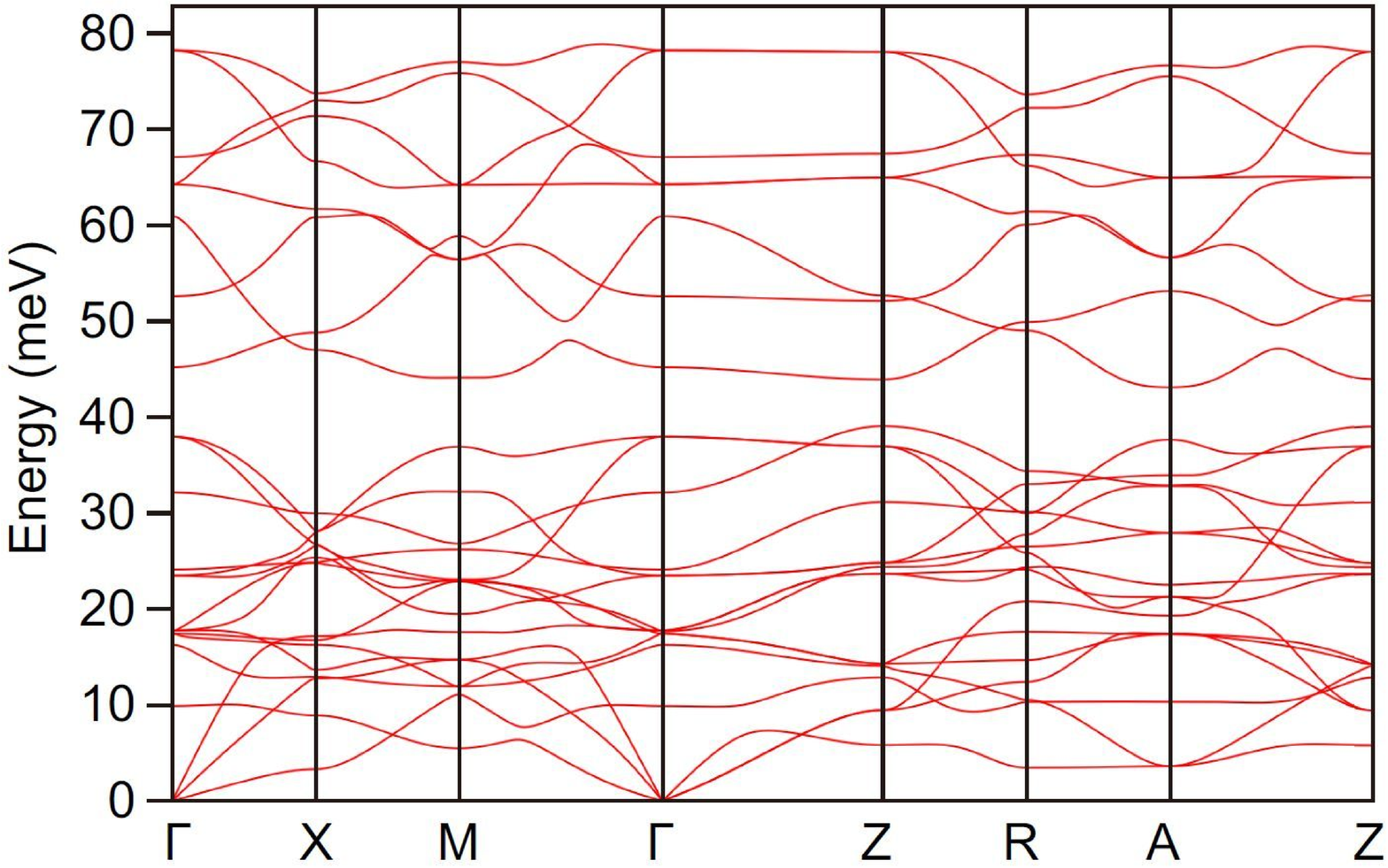}
\caption{
Phonon dispersion of SrCu$_3$O$_4$ obtained by first-principles 
calculation using the lattice constants of the thin film 
($a = 5.42$~\AA\, and $c = 3.53$~\AA).
}
\label{fig.s7}
\end{center}
\end{figure}

The phonon dispersion of SrCu$_3$O$_4$ calculated using the experimentally
observed lattice constants of $a = 5.42$~\AA\, and $c = 3.53$~\AA\, 
is shown in Figure~\ref{fig.s7}. 
No imaginary modes are present, suggesting that SrCu$_3$O$_4$ is 
dynamically stable. This result is consistent with the experimental observation
of the multiple domain formation for thick films of SrCu$_3$O$_4$ 
(Figures~\ref{fig.s5} and \ref{fig.s6}); that is, SrCu$_3$O$_4$ is 
considered thermodynamically stable.
%

\providecommand{\latin}[1]{#1}
\makeatletter
\providecommand{\doi}
  {\begingroup\let\do\@makeother\dospecials
  \catcode`\{=1 \catcode`\}=2 \doi@aux}
\providecommand{\doi@aux}[1]{\endgroup\texttt{#1}}
\makeatother
\providecommand*\mcitethebibliography{\thebibliography}
\csname @ifundefined\endcsname{endmcitethebibliography}
  {\let\endmcitethebibliography\endthebibliography}{}

\end{document}